\documentstyle[preprint,aps]{revtex}
\begin{document}
\draft
\tighten
\title{Neutrino induced transitions between the ground states of the A=12 
triad}
\author{J. Engel$^1$, E. Kolbe$^2$, K. Langanke$^3$,  and P. Vogel$^4$}
\address{
$^1$ Department of Physics, University of North Carolina, Chapel Hill, NC 
27599 \\
$^2$ Physikalishes Institut, Universit\"{a}t Basel, Switzerland \\
$^3$ W.K. Kellogg Radiation Lab., California Institute of Technology, 
Pasadena, CA 91125 \\
$^4$ Physics Dep., California Institute of Technology,
Pasadena, California 91125 }
\date{\today}
\maketitle

\begin{abstract}

Neutrino induced reactions on $^{12}$C, an ingredient of liquid
scintillators, have been studied in several experiments.  We show that for
currently available neutrino energies, $E_{\nu} \le$ 300 MeV, calculated
exclusive cross sections $^{12}$C$_{gs}(\nu,l)$$^{12}$N$_{gs}$ for both muon
and electron neutrinos are essentially model independent, provided the
calculations simultaneously describe the rates of several other reactions
involving the same states or their isobar analogs.  The calculations agree
well with the measured cross sections, which can be therefore used to check
the normalization of the incident neutrino spectrum and the efficiency of the
detector.

\end{abstract}


A solid theoretical understanding of cross sections in neutrino-induced
reactions on light nuclei is becoming a necessity, in particular for
$^{12}$C, an ingredient of liquid scintillators, and for $^{16}$O, the basic
component of water \v{C}erenkov detectors.  Detectors containing these
isotopes are used to measure fluxes of atmospheric and supernova neutrinos,
and in searches for neutrino oscillations.  These tasks to a large extent
require knowledge of the corresponding cross sections, which often have not
been measured.  When experimental cross sections are available it is
therefore important to compare them with calculations.

For $^{12}$C a number of experimental results exist.  These include
measurements of charged-current reactions induced by both
electron- \cite{Karmen} and muon-neutrinos \cite{Koetke,LSND,Louis},
exciting both the ground and continuum states in $^{12}$N.  The inclusive
cross section for $^{12}$C($\nu_e,e$)$^{12}$N$^{*}$, measured by Karmen
\cite{Karmen}, LAMPF (with large errors)\cite{Krakauer}, and recently by the
LSND collaboration \cite{Louis}, agree well with calculations.  [A previously 
reported measurement of both exclusive and inclusive
neutrino cross sections on $^{12}$C \cite{Koetke} appears to be
inconsistently large \cite{LSND}; we will not consider that
measurement here.]  By contrast, there is a disturbing discrepancy between
calculations\cite{Kolbe94,Kolbe95,Oset} and the LSND value of the cross
section for the inclusive reaction $^{12}$C($\nu_\mu,\mu$)$^{12}$N$^{*}$,
which uses higher energy neutrinos from pion-decay in flight
\cite{LSND,Louis,thesis}.  The disagreement is disturbing\footnote{ The
disagreement is not universal.  In \cite{Mintz} a cross section that agrees
with LSND is obtained using the elementary particle method.  We discuss the
applicability of the approximations used in \cite{Mintz} elsewhere
\cite{aMintz}.}  in light of the apparent simplicity of the reaction
and in view of the fact that parameter-free calculations, such as
those in \cite{Kolbe94,Kolbe95} describe well other weak processes
which are governed by the same weak-current nuclear matrix elements. 
Here we examine whether similar problems 
affect our understanding of the exclusive
reactions $^{12}$C($\nu_e,e$)$^{12}$N$_{gs}$\cite{Karmen,LSND} and
$^{12}$C($\nu_\mu,\mu$)$^{12}$N$_{gs}$\cite{LSND,Louis,thesis}.  The
exclusive process is a useful monitor of the neutrino flux.  If calculations
are reliable and reproduce experiment, then the normalization of the
experimental neutrino flux must have been correctly modeled.

A calculation of the exclusive cross section can be tested by computing rates
of related processes and comparing to data.  The $1^+~ T=1$  ground
state in $^{12}$N is the analog of the 15.11 MeV state in $^{12}$C and
of the ground state of $^{12}$B.  This allows us to
use $\beta^+$ decay from $^{12}$N back to the $0^+~ T=0$ ground state of
$^{12}$C, muon capture from $^{12}$C to the ground state of
$^{12}$B, ineleastic electron scattering to the 15.11 MeV state
in $^{12}$C, and $M1$ decay from that state to the ground state to
calibrate elements of the calculation.  In what follows we calculate the
exclusive neutrino cross sections in several ways to see whether when all
the above data are reproduced, the different models can produce significantly
different predictions.

Our first approach is a restricted shell-model calculation.  Describing all
the above reactions is straightforward in the shell model as long as we can
neglect contributions of two-body operators (i.e.\ meson exchange).  The
reduced matrix element of an arbitrary operator \^{O} is then given by
\begin{equation}
\langle 1^+,1 || \hat{O} || 0^+,0 \rangle = \sum_{j,j'}
\langle j || \hat{O} || j' \rangle OBD(j,j') ~,
\end{equation}
where the one-body transition densities are defined by
\begin{equation}
 OBD(j,j') = \langle 1^+,1 || [a_j^{\dagger} \tilde{a}_j]^{J^{\pi},T=1^+,1} 
|| 0^+,0 \rangle ~.
 \end{equation}
If we further assume that all structure in the low-lying states with
$J^{\pi}=1^+,T=1$ is generated by the eight ``valence" nucleons moving in the
$p$-shell, there are only 4 one-body densities and 4 single-particle matrix
elements $\langle j || \hat{O} || j' \rangle$, which contain all the
momentum-transfer dependence and that are simple to calculate.  Futhermore, the
Gamow-Teller (GT) matrix element for the $\beta^+$ decay of $^{12}$N, the
$M1$ radiative width of the state at 15.11 MeV in $^{12}$C, and the form
factor for the excitation of this state in electron scattering \cite{elec}
depend only on {\em three} one-body densities; they are independent of
the combination $OBD(p_{3/2},p_{1/2}) + OBD(p_{1/2},p_{3/2})$.  (See the
review \cite{Peccei} for a general discussion.)

The most straightforward way of obtaining the one-body densities is by
diagonalizing a thoroughly tested residual interaction, such as 
the one given in Ref.
\cite{WB}.  It is well known, however, that the resulting $p$-shell one-body
densities do not describe the reactions above very well.  Typically, the
calculated GT strength is too large (the origin of ``GT quenching'') and the
electron scattering form factor is too high
in the first lobe.  In a number of papers\cite{Haxton,Donnelly,Brady,Nimai}
over the last twenty plus years, however, it has been shown that one can
modify the one-body densities (ad hoc) in such a way that the three
experiments are correctly reproduced, with the form factor adequately
described up to the first minimum at $|\vec q| \approx 1.5$ fm$^{-1}$.  This
is the approach we follow here, adjusting the densities to reproduce the data.
The so-far undetermined fourth combination of densities can be fixed (with
some uncertainty) by the muon capture rate to the ground state of $^{12}$B
\cite{Peccei}.
This process is the only one that tests the momentum dependence of the axial 
current, since it takes place at $q^2 = -0.74m_{\mu}^2$.  For our
analysis we take an averaged experimental rate, $\omega(1^+) = 6200\pm200$
s$^{-1}$ \cite{capt}, and use the Goldberger-Treiman relation for the
induced-pseudoscalar coupling constant.

Line 2 of Table I contains the resulting one-body densities, adjusted here to
reproduce all the data discussed above.  Lines 3-5 contain
one-body densities used by other authors, and constrained to 
different subsets of the above data. To achieve an overall agreement
with all the data, we renormalize, in addition, either the weak axial 
form factor $F_A$ or the magnetic form factor $F_M$. 
Haxton\cite{Haxton} (line 3) required that the $\beta^-$ decay of $^{12}$B, 
rather than $^{12}$N (but not
the muon capture), be well described; because of isospin violation in the $ft$
values, his densities require a renormalization factor of
0.873 for the axial current form factor $F_A$ to fit the $^{12}$N decay.  
The entries in line 4 are
based on the extreme single particle model \cite{Donnelly}, in which $^{12}$C
is represented as a closed $p_{3/2}$ subshell, and $^{12}$N or $^{12}$B have
one nucleon in $p_{1/2}$ subshell.  Here a renormalization of 0.414 is
required for $F_A$, and 0.484 for the magnetic form factor $F_M$.  Finally,
line 5 contains the one-body densities of Ref.\ \cite{Brady}, based on a
minimally modified Cohen-Kurath interaction.  They require only a small $F_A$
renormalization of 0.925.  In all cases harmonic oscillator wave functions
are used, with slightly different values 
of the oscillator parameter $b$  taken from the original references.  
The table clearly illustrates that the
extraction of one-body densities from data is not a unique procedure; it
depends on other assumptions about wave functions, etc.  We discuss the
effects of these differences on the exclusive neutrino cross sections
shortly.

While two-body effects, i.e.\ meson exchange corrections, are expected to be
relatively small (5-10 \%) for the momentum transfer considered
here\cite{Umino}, configurations beyond the $p$ shell might explain the need
for a drastic renormalization of the one-body densities produced by a
reasonable $p$-shell Hamiltonian.  We therefore calculate the rates of all
the reactions above, including exclusive neutrino capture, in the Random
Phase Approximation (RPA), which does include multishell correlations, while
treating the configuration mixing within the $p$ shell only crudely.  The
same method, extended to the continuum, has been used in the calculation of the
neutrino charged and neutral current inclusive scattering
\cite{Kolbe94,Kolbe95,Kolbe92}, (disagreeing disturbingly with one
experiment, as noted above) and muon capture \cite{Kolbe94b}.

To compensate for the crude description of the $p$-shell dynamics we use an
overall quenching factor of 0.258 by which we multiply the rates, respectively
the cross sections of all processes
under consideration, for all momentum transfers.  With this multiplicative factor the
$\beta^+$ decay, $M1$ width, electron scattering form factor and partial muon
capture rate are all adequately described, and the exclusive cross sections,
discussed below, are readily calculated.
(It has been known for some time that the RPA is capable of describing the shape
of the $(e,e')$ form factor for the 15.11 MeV state \cite{Suzuki}. )

Our third approach is the ``elementary-particle treatment" (EPT).  Instead of
describing nuclei in terms of nucleons or other constituents, the EPT
considers them elementary and describes transition matrix elements in
terms of {\it nuclear} form factors deduced from experimental data and
constrained by transformation properties.  The EPT was implemented in the
$A=12$ system in Refs.\ \cite{Fukugita,Kubodera,Kubodera2} for neutrino
energies up to $E_{\nu}$ = 100 MeV.  Here we extend the approach to the
higher neutrino energies relevant to the LSND decay-in-flight $\nu_{\mu}$'s
by appropriately including the lepton mass, which was ignored in Ref.\
\cite{Fukugita}, in the kinematics.  A nonzero lepton mass requires, in turn,
that the induced pseudoscalar
term, also neglected in Ref.\ \cite{Fukugita}, must be included as well.  
Here we used a modified form of Eq.\ (16) in
Ref.\ \cite{Fukugita}, kindly furnished to us by Professor Kubodera and for the 
$F_P(Q^2)$ we used the simple ansatz:
\begin{equation}
F_P(Q^2) = - \frac{m_{\pi}^2}{Q^2} F_A(0)
\left[ F_A(Q^2) - \frac{m_{\pi}^2}{Q^2 + m_{\pi}^2}\frac{1}{1 + Q^2/\xi^2} 
\right]~,
\end{equation}
where, as usual, $Q^2 = -q^2$, and the empirical parameter $\xi$ has been 
fixed from muon capture.

We turn now to the evaluation of the exclusive cross sections in the three
approaches.  Within the shell model and the RPA the cross section is easily
evaluated once the one-body densities and free-nucleon form factors (for
which we use standard values) are specified; the general prescription can be
found, e.g., in Ref.  \cite{Walecka}.  In the EPT the evaluation is even
simpler since the nucleus is elementary. One complication often not
considered is related to the Coulomb interaction of the outgoing charged
lepton.  The usual treatment, as in nuclear $\beta$ decay, involves the Fermi
function $F(Z,E)$, which is the ratio of the Coulomb continuum $s$-wave and
the corresponding free $s$-wave wave.  This approximation is valid, however,
only for lepton momenta $pR \le 1$, where $R$ is the nuclear radius.  For
$^{12}$C it can be used for electrons with up to
about 50 MeV of kinetic energy, but it is justified only for muons of 10 MeV 
or less of kinetic energy.  As energy increases, $F(Z,E)$ approaches a
constant value, $\sim$ 1.17  ($Z=7$)  for $\nu$ reactions on $^{12}$C,
and $\sim$ 0.90  ($Z=5$) for $\bar{\nu}$-induced reactions
on $^{12}$C. On the other hand,
it is intuitively clear that as the lepton energy  
becomes much larger than the Coulomb 
energy, the Coulomb correction
should approach unity. 
In order to keep things simple, and since the 
correction is a relatively minor one,
we scale at higher energies the lepton cross section by the ratio
\begin{equation}
\frac{p_{eff} (E + \langle V \rangle)}{pE} ~,{\rm where} ~
p_{eff} = p(1 + \frac{\langle V \rangle}{E}), ~ \langle V \rangle = \pm
\frac{3Z\alpha}{2R} ~,
\end{equation}
and $\langle V \rangle$ represents
the average Coulomb potential. The two approximations are smoothly connected.

The results appear in Table II.  For the $\nu_e$-induced reaction with
neutrinos from the muon decay-at-rest \cite{Karmen} the agreement between the
experimental and calculated exclusive cross section is perfect in all the
models.  This cross section, corresponding to an average momentum transfer of
only about 50 MeV, appears totally fixed by the requirements we impose on
each of the calculations.  For the $\nu_{\mu}$-induced reaction the average
momentum transfer is about 150 MeV.  But even in this case the different variants of
the shell model, the RPA and the EPT give quite similar cross sections.  Again,
the agreement with the LSND experimental value is good.

In Figures 1 and 2 we show the exclusive cross sections for
$^{12}$C($\nu_e,e$)$^{12}$N$_{gs}$ and
$^{12}$C($\nu_\mu,\mu$)$^{12}$N$_{gs}$, respectively, as a function of
neutrino energy.  For $\nu_e$'s the cross section is essentially the same in
all the approaches we consider.  The agreement between the various models is
also quite good for $\nu_{\mu}$'s.  Moreover, after a very steep rise from
the threshold, this cross section quickly saturates and becomes more or less
independent on the neutrino energy.  Thus, the exclusive cross section is
simply proportional to the {\it total} number of $\nu_{\mu}$'s above
threshold. However,  it is essentially independent of their energy distribution.

For completeness, Table III lists the cross sections for neutral-current
excitation of the 15.11 MeV, $1^+, T=1$ state in $^{12}$C, which was measured 
by Karmen \cite{Karmen} for neutrinos from decay at
rest.  Once again all the models agree with one another and the data.  The
neutral-current cross section for $\nu_{\mu}$'s from decay-in-flight and for
$\bar{\nu}_{\mu}$'s are probably not measurable, but unsurprisingly the
calculations continue to agree with one another.

In conclusion, we have shown that the exclusive cross sections for neutrino
energies available now are calculable in a variety of ways, with results that
are nearly identical, and agree very well with the data.  
These processes can therefore be used as a
check of the neutrino flux normalization, and detector efficiency
and indicate that the
discrepancy between the measured and calculated inclusive
$^{12}$C($\nu_\mu,\mu$)$^{12}$N$^{*}$ cross sections is not caused by an
underestimate of the neutrino flux normalization.  The source the
disagreement must be found elsewhere.

We would like to thank Professor Kubodera for his help with the EPT.  
We also appreciate numerous discussions with the members of
LSND collaboration. We were
supported in part by the U.S.\ Department of Energy under grants
DE-FG05-94ER40827 and DE-FG03-88ER-40397, by the U.S.  National Science
Foundation under grants PHY94-12818 and PHY94-20470, and by the Swiss
Nationalfonds.

\begin{table} [htb]
\caption {\protect Fitted one-body densities  $OBD(j,j')$.
The parameter $b$ is the oscillator length. The range in line 2, column 4,
reflects the uncertainty in the muon capture rate; in other
lines only the most probable capture rate was used.}
\label{tab:obd}
\begin{center}
\begin{tabular}{l|lllll} 
$j,j'$ & 1/2,1/2 & 1/2,3/2-3/2,1/2 & 1/2,3/2+3/2,1/2 & 3/2,3/2 & $b$ (fm) \\
\hline
present & -0.113 & 0.106 & 0.666$\pm$0.4 & 0.24 & 1.67 \\
Ref. \cite{Haxton}~~ & -0.111 & 0.337 & 0.875 & 0.086 & 1.76 \\
Ref. \cite{Donnelly}~~ & ~0.0 & 1.0 & 1.0 & 0.0 & 1.77 \\
Ref. \cite{Brady}~~ & -0.27 & 0.318 & 1.03 & 0.116 & 1.82 \\
\end{tabular}
\end{center}
\end{table}

\begin{table} [htb]
\caption {\protect Comparison of calculated and measured cross sections, in 
units of $10^{-42}$cm$^{-2}$ and averaged over
the corresponding neutrino spectra, for the neutrino induced transitions 
$^{12}$C$_{gs} \rightarrow ^{12}$N$_{gs}$ and $^{12}$C$_{gs} \rightarrow 
^{12}$B$_{gs}$. For the decay at rest the $\nu_e$ spectrum is normalized
from $E_{\nu}$ = 0 while for the decay in flight the $\nu_{\mu}$ and $\bar{\nu}_{\mu}$
spectra are normalized from the corresponding threshold.
The cross section for $\bar{\nu}_e$ is not quoted since the decay-at-rest 
neutrino source does not contain any $\bar{\nu}_e$'s.}
\label{tab:gs}
\begin{center}
\begin{tabular}{lccc} 
\hline
  & $^{12}$C($\nu_e,e^-)^{12}$N$_{gs}$  & 
$^{12}$C($\nu_{\mu},\mu^-)^{12}$N$_{gs}$  &
 $^{12}$C($\bar{\nu}_{\mu},\mu^+)^{12}$B$_{gs}$  \\
  &  decay at rest   & decay in flight  & decay in flight \\
\hline  \\
\bigskip
experiment & 8.9$\pm 0.6 \pm 0.75$ \cite{Karmen}&  - & - \\
\bigskip
experiment & 9.1$\pm 0.4 \pm 0.9$ \cite{Louis} & 64$ \pm 10 \pm 10 $
\cite{Louis} & - \\
\bigskip
OBD of this work & 9.1 $\pm$ 0.1 & 63.5$^{\dagger}$ $ \pm $5  & 24 \\
\bigskip
OBD of Ref.\cite{Haxton} & 8.8 & 60.4 & 23 \\
\bigskip
OBD of Ref.\cite{Donnelly}  & 9.4 & 65.4 & 22.6 \\
\bigskip
OBD of Ref \cite{Brady} & 9.4 & 62.3 & 23.7 \\
\bigskip
CRPA \cite{Kolbe94,Kolbe95} & 8.9 & 63.0 & 26 \\
\bigskip
EPT$^{\dagger\dagger}$ \cite{Fukugita} & 9.2 & 62.9 &  21.5\\
\end{tabular}
\end{center}
$^{\dagger}$  The uncertainty reflects the range corresponding to the 
uncertainty
in the muon capture rate. A similar range is presumably valid for the other 
approaches in lines 4-6. 

$^{\dagger\dagger}$ Extended to muon neutrinos, K. Kubodera, private 
communication.
\end{table}

\begin{table} [htb]
\caption {\protect Comparison of calculated and measured cross sections, in 
units of $10^{-42}$cm$^{-2}$ and averaged over
the corresponding neutrino spectra, for the neutrino-induced neutral-current 
transitions $^{12}$C$_{gs} (\nu,\nu')
^{12}$C (15.11). In column 2 the contributions from $\nu_e$ and $\bar{\nu}_{\mu}$
are added. All spectra are normalized ``per neutrino'' i.e. from $E_{\nu}$=0.}
\label{tab:nc}
\begin{center}
\begin{tabular}{lccc} 
\hline
  & $^{12}$C($\nu_e/\bar{\nu}_{\mu}, \nu_e' /\bar{\nu}_{\mu}')^{12}$C(15.11)  
& 
$^{12}$C($\nu_{\mu},\nu_{\mu}')^{12}$C (15.11)  &
 $^{12}$C($\bar{\nu}_{\mu}, \bar{\nu}_{\mu}') ^{12}$C (15.11)  \\
  &  decay at rest   & decay in flight  & decay in flight \\
\hline  \\
\bigskip
experiment & 11$\pm 1.0 \pm 0.9$ \cite{Karmen}& - & - \\
\bigskip
OBD of this work & 9.3 & 22.3  & 15.2 \\
\bigskip
OBD of Ref.\cite{Haxton} & 9.3 & 21.9 & 15.3 \\
\bigskip
OBD of Ref.\cite{Donnelly}  & 9.4 & 23.7 & 16.4 \\
\bigskip
OBD of Ref \cite{Brady} & 9.6 & 22.9 & 15.9 \\
\bigskip
CRPA \cite{Kolbe94,Kolbe95} & 10.5 & 27.2 & 17.5 \\
\bigskip
EPT \cite{Fukugita} & 9.4 & 24.3 &  14.5 \\
\end{tabular}
\end{center}
\end{table}

\begin{figure}
\caption[Figure 1]
{Cross section for $^{12}$C($\nu_e,e$)$^{12}$N$_{gs}$ as a function
of the $\nu_e$ energy, in units of $10^{-42}$cm$^2$. The full line is for the
shell model with one-body densities from line 2, Table I,
the dashed line
is the modified Cohen-Kurath one body densities from line 5, Table I \cite{Brady}
(the curves for the other shell model variants are very
similar and are not shown), 
the dot-and-dashed line is for the RPA \cite{Kolbe95}, and the dotted line is for 
the EPT.}
\label{f:elec}
\end{figure}

\begin{figure}
\label{f:muon}
\caption[Figure 2]
{Cross section for $^{12}$C$\nu_\mu,\mu$)$^{12}$N$_{gs}$ 
as a function of the $\nu_{\mu}$ energy, in units of $10^{-42}$cm$^2$. 
The notation is as in Fig. \ref{f:elec}.}
\end{figure}

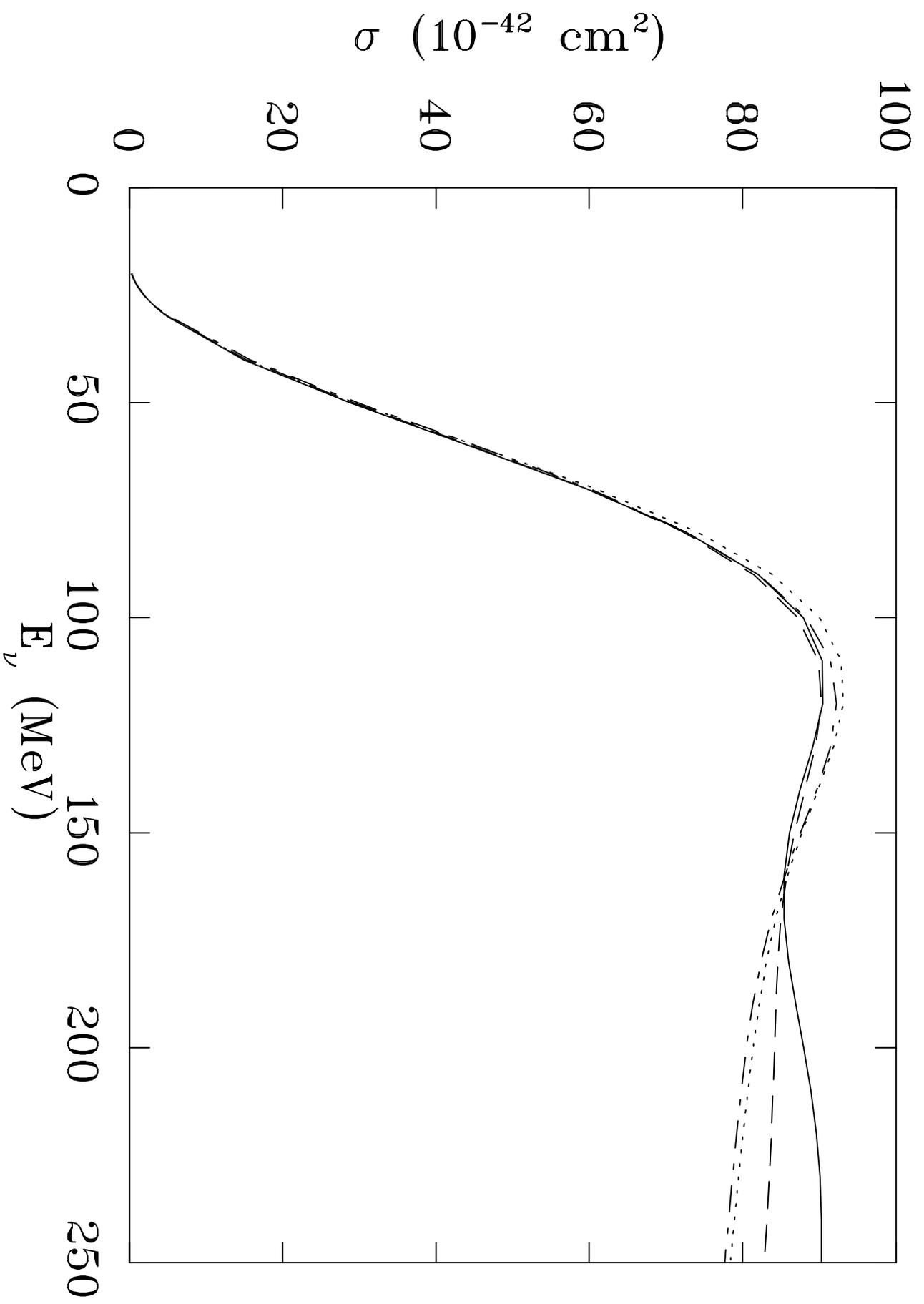
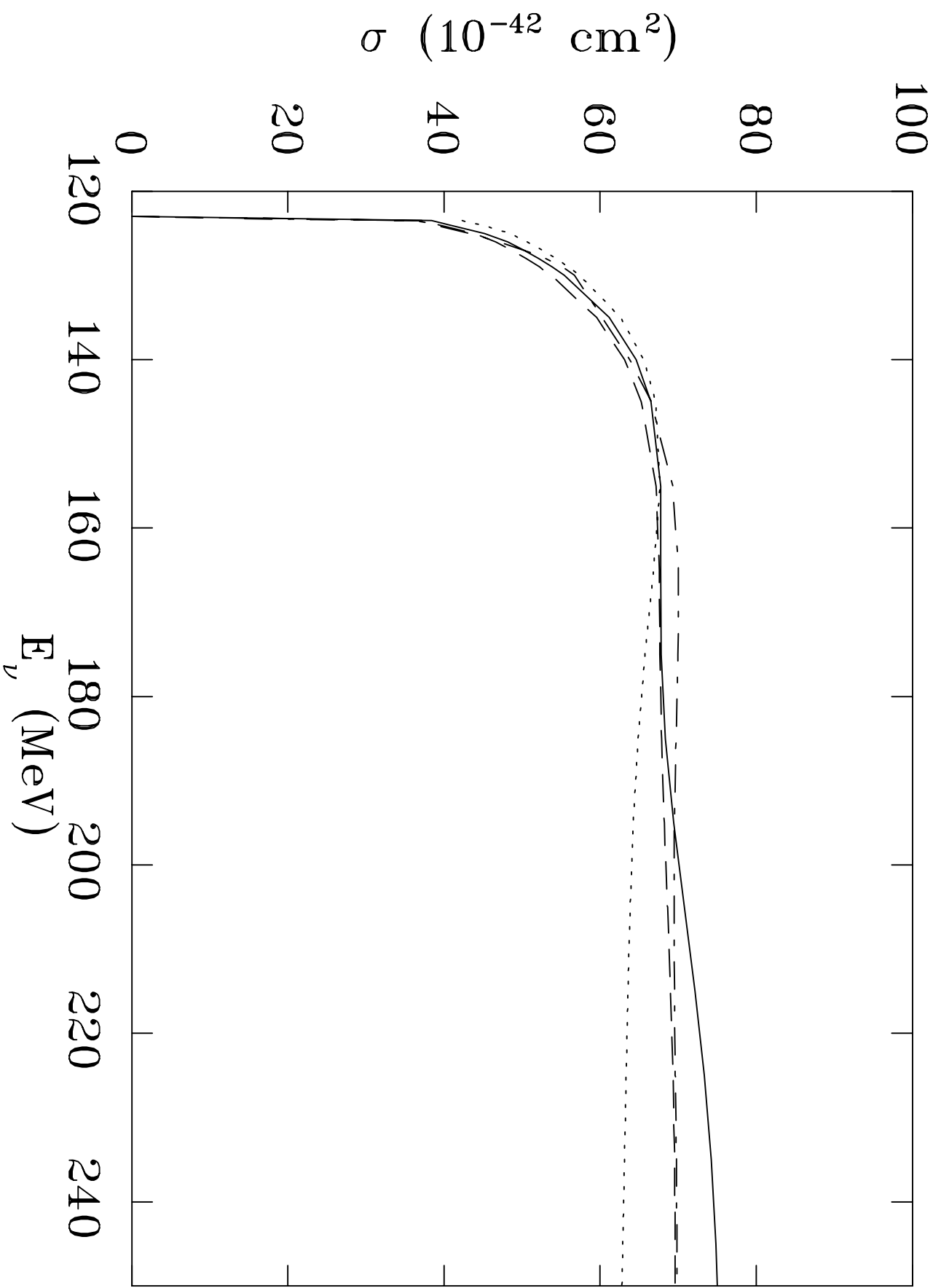
\end{document}